\documentclass[a4paper,12pt]{article}
\usepackage{geometry}
\usepackage{lscape}
\usepackage{array}
\usepackage{tikz}
\usepackage{tikz-qtree}

\usepackage{graphicx}
\usepackage{amsmath}
\usepackage{bbm}
\usepackage{cite}
\usepackage[sort&compress,numbers]{natbib}
\usepackage{amsfonts}
\usepackage{amssymb}
\usepackage{amsmath}
\usepackage{latexsym}
\usepackage{color}
\usepackage{url}
\usepackage{ntheorem}
\usepackage{braket}
\usepackage[normalem]{ulem}
\usepackage{enumerate}
\usepackage[colorlinks=true]{hyperref}

\newcommand{\ptcignore}[1]{}

\newcommand{\bredA}{\red{\bar A}}
\newcommand{\bdelta}{\red{\bar \delta}}

\newcommand{\redc}{\color{red} }

\newcommand{\redA}{\red{A}}

\newcommand{\scaleB}{\blue{  B}}
\newcommand{\Binf}{\blue{\bar B}}

\newcommand{\ptcheck}[1]{\ptc{checked on #1}}

\newcommand\ben{\begin{enumerate}}
\newcommand\een{\end{enumerate}}
\newcommand\bit{\begin{itemize}}
\newcommand\eit{\end{itemize}}

\newcommand{\blue}[1]{{\color{blue}#1}}
\newcommand{\red}[1]{{\color{red}#1}}

\newcounter{mnotecount}[section]

\renewcommand{\themnotecount}{\thesection.\arabic{mnotecount}}

\newcommand{\mnote}[1]%{}
{\protect{\stepcounter{mnotecount}}$^{\mbox{\footnotesize
$%\!\!\!\!\!\!\,
\bullet$\themnotecount}}$ \marginpar{%\color{red}%
\raggedright\tiny\em
$\!\!\!\!\!\!\,\bullet$\themnotecount: #1} }

\newcommand{\jlcax}[1]{}
\newcommand{\eean}{\nonumber\end{eqnarray}}

\newcommand{\kk}[1]{}%{\mnote{{\bf If we consider the KK case:} #1}}

\newcommand{\mcH}{{\mycal H}}

\newcommand{\beq}{\begin{equation}}

\newcommand{\FS}       %{F_1} %
                  {F}
\newcommand{\HS} %{F_2}
       {H_{\mbox{\scriptsize volume}}}

{\ptc{this should be removed in the oberwolfach version}}%

\newcommand{\eeal}[1]{\label{#1}\end{eqnarray}}
\newcommand{\bed}{\begin{deqarr}}
\newcommand{\eed}{\end{deqarr}}
\newcommand{\bedl}[1]{\begin{deqarr}\label{#1}}
\newcommand{\eedl}[2]{\arrlabel{#1}\label{#2}\end{deqarr}}

\newcommand{\bel}[1]{\begin{equation}\label{#1}}
\newcommand{\bea}{\begin{eqnarray}}
\newcommand{\bean}{\begin{eqnarray}\nonumber}
\newcommand{\beal}[1]{\begin{eqnarray}\label{#1}}
\newcommand{\eea}{\end{eqnarray}}

\newcommand{\be}{\begin{equation}}
\newcommand{\eeq}{\end{equation}}
\newcommand{\ee}{\end{equation}}
\newcommand{\beqa}{\begin{eqnarray}}
\newcommand{\eeqa}{\end{eqnarray}}
\newcommand{\beqan}{\begin{eqnarray*}}
\newcommand{\eeqan}{\end{eqnarray*}}
\newcommand{\ba}{\begin{array}}
\newcommand{\ea}{\end{array}}

\newcommand{\warn}[1]%{}%{}
{\protect{\stepcounter{mnotecount}}$^{\mbox{\footnotesize
$%\!\!\!\!\!\!\,
\bullet$\themnotecount}}$ \marginpar{%\color{red}%
\raggedright\tiny\em
$\!\!\!\!\!\!\,\bullet$\themnotecount: {\bf Warning:} #1} }

\newcommand{\R}{\mathbb R}

\newcommand{\ptc}[1]{\mnote{{\bf ptc:}#1}}

\newcommand{\beqar}{\begin{deqarr}}
\newcommand{\eeqar}{\end{deqarr}}

\newcommand{\beaa}{\begin{eqnarray*}}
\newcommand{\eeaa}{\end{eqnarray*}}

\DeclareFontFamily{OT1}{rsfs}{}
\DeclareFontShape{OT1}{rsfs}{CGNPm}{n}{ <-7> rsfs5 <7-10> rsfs7 <10-> rsfs10}{}
\DeclareMathAlphabet{\mycal}{OT1}{rsfs}{CGNPm}{n}

{\catcode `\@=11 \global\let\AddToReset=\@addtoreset}
\AddToReset{equation}{section}

{\catcode `\@=11 \global\let\AddToReset=\@addtoreset}
\AddToReset{figure}{section}

{\catcode `\@=11 \global\let\AddToReset=\@addtoreset}
\AddToReset{table}{section}

\renewcommand{\ptcheck}[1]{} %toremove colours
\renewcommand{\red}[1]{#1}%toremove colours
\renewcommand{\redc}{}%toremove colours
\renewcommand{\blue}[1]{#1}%toremove colours
\begin{document}
\title{Static vacuum metrics in $(4+1)$ dimensions with $\Lambda < 0$ 
and  squashed conformal infinity 
}

\author{Piotr T. Chru\'{s}ciel\thanks{Beijing Institute of Mathematical Sciences and Applications, Huairou, and Center for Theoretical Physics of the Polish Academy of Sciences, Warsaw} \thanks{
{\sc Email} \protect\url{pchrusciel@cft.edu.pl}, {\sc URL} \protect\url{homepage.univie.ac.at/piotr.chrusciel}}
\\  
Jiaxin Song\thanks{School of Physics, Nankai University, 300071 Tianjin, China}
\\ 
{Raphaela Wutte}\thanks{Mathematical Sciences and STAG Research Centre, University of
Southampton, %Highfield, SO17 1BJ Southampton, 
United Kingdom} 
\thanks{
{\sc Email} \protect\url{rwutte@hep.itp.tuwien.ac.at} }
\\ 
{Maciej Maliborski}\thanks{TU Wien,  Institut für Analysis und Scientific Computing,
%Wiedner Hauptstraße 8–10
%1040 
Vienna, Austria}
\\ 
{Klaus Kr\"oncke\thanks{Department of Mathematics, KTH Royal Institute of Technology, 
%10044 
Stockholm, Sweden}}
}
\maketitle

\begin{abstract}
We construct numerically two families of  static five-dimensional Lorentzian metrics, solutions of vacuum Einstein equations with a negative 
cosmological constant, one with and one without a black hole region, with a metric at sections of conformal infinity which is conformal to any squashed three-dimensional sphere.
\end{abstract}

\tableofcontents
%----------------------------------------------------------------------------%

\ptcignore{see also a forgotten problem numerics.tex, which becomes interesting in view of the Brendle-Hung proof of the Horowitz-Myers conjecture}

 \ptcignore{globally reread on 13VII26}
\section{Introduction}

In~\cite{ChDelayKlingerBH,ACD2} large classes of static solutions of the vacuum Einstein equations with a negative cosmological constant have been constructed near Birmingham-Kottler  solutions. The aim of this note is to point out existence of  a family of five-dimensional such metrics with spherical infinity which  are \emph{far away} from the Birmingham-Kottler ones.

The starting point of the analysis here are the results in~\cite{ChDelayKlingerBH,ACD2}, where it is shown that near every static vacuum asymptotically locally hyperbolic (ALH) metric $g$, with the property that the Lichnerowicz operator of the Wick-rotated counterpart of $g$ has no kernel, there exists a large family of static metrics which are uniquely determined by a set of free data at conformal infinity. The kernel property is satisfied by many, and most likely almost all~\cite{ChDelayKlingerNonDegenerate,KlingerBirkhoff}, Birmingham-Kottler metrics. The uniqueness part of the above statement guarantees that symmetries of the data at infinity descend to the interior.

 For the usual  Birmingham-Kottler metrics which are positively-curved at infinity, the space-part of conformal infinity is a round sphere 
  which, in spacetime dimension five, belongs to the two-parameter family of $SO(3)\times U(1)$-symmetric squashed spheres (cf., e.g., \cite{BizonRostworowskiAds5}). 
The results just mentioned, together with the analysis in~\cite{ChDelayKlingerNonDegenerate,KlingerBirkhoff} of the relevant Lichnerowicz operator, 
 guarantee that there exist non-trivial static solutions with these symmetries near the Anti-de Sitter spacetime, and near many (and, as already mentioned, probably almost all) remaining Birmingham-Kottler metrics; the former will be conformally compact without event horizons, while the latter will have black hole regions.  The question then arises of existence of such solutions which are far from the spherically-symmetric ones.   
 In this note we provide numerical evidence that, for every metric at infinity which is  conformal to an ultrastatic metric with space sections which are  squashed three-dimensional spheres,   there exists a unique, within the family, static vacuum fill-in metric  without horizons, as well as  a one-parameter family of black hole metrics with non-degenerate horizons.
  \ptcignore{is there a squashed version of negatively curved infinity?}

A plot of  the holographic mass~\cite{Skenderis:2000in,SkenderisCheng}, say $m_{\textrm{hol}}  $, for horizon-less solutions, 
as a function of the asymptotic value $\Binf_0$ of  the squashing function $B$ (cf.~\eqref{eq:25.07.23_01} and \eqref{12VIII25.1b} below), can be found   in Figure~\ref{F13VII25.1}. 
\begin{figure}
    \centering%
%    \includegraphics[width=0.4\textwidth]{A4l.pdf}\hfill
%  \includegraphics[width=0.4\textwidth]{A44.pdf}
%    \vspace{1em} 
%%    \includegraphics[width=0.4\textwidth]{A4.pdf}\hfill
%%    \includegraphics[width=0.4\textwidth]{A43.pdf}
    \includegraphics[width=0.49\textwidth]{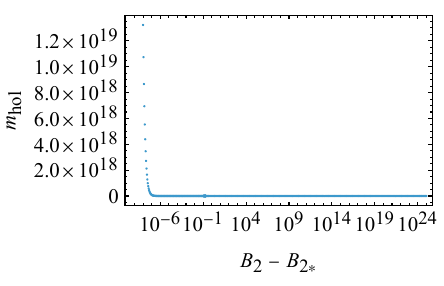}
%    \caption{The coefficient $A_4$ for solutions without horizons as a function of $B_2$ (left plot) and $\Binf_0$ (right plot). 
%%    The bottow row is a zoom into the region where $A_4$ attains a maximum.
%    \\
%    \\
%    }\label{F1V26.1}%
%%\end{figure}
%% %
\hspace{-0.4cm}
%%\begin{figure}
%	\centering
%	  \includegraphics[width=0.4\textwidth]{mb2}\hfill
    \includegraphics[width=0.49\textwidth]{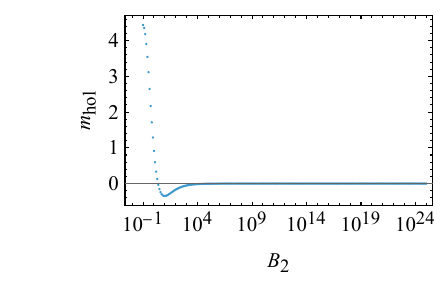}
    \vspace{1em} 
	\caption{Holographic mass of solutions without horizons. 
 The mass goes monotonically to infinity when $B_2\searrow B_{2*}$, which (numerically) appears to correspond to $\Binf_0\to -\infty$.}   
\label{F13VII25.1} 
\end{figure}
Recall that static vacuum metrics in $(3+1)$ dimensions have negative holographic mass~\cite{HicklingWiseman}. Here we find solutions with both positive and negative mass, with a phase transition at $\Binf_0\approx 0.726$. 
In fact, from the perspective of the mass parameter, we find exactly one solution without horizons for all $m_{\textrm{hol}}\ge 0 $ and for 
$$m_{\textrm{hol}}=m_*\approx -0.34764
\,,
$$
no solutions with $m_{\textrm{hol}} < m_* $, and exactly two solutions with distinct $\Binf_0$ for $m_* <m_{\textrm{hol}} <0 $.

\section{The metrics}
\label{sec:themetrics}
We consider $SO(3)
\times U(1)$-symmetric metrics of the form 
\begin{align}
g &= %
%  -\tilde A e^{-2\tilde\delta}dt^{2} + A^{-1}dx^{2} + \frac{1}{4}r^2\left(e^{2B}\left(\sigma_{1}^{2}
% +\sigma_{2}^{2}\right)+e^{-4B}\sigma_{3}^{2}\right) 
% \\
%  & =
 - \tilde{a}(r) dt^2 + \tilde{b}(r) dr^2 + \frac{1}{4} r^2
\left(e^{\tilde{c}(r)} (\sigma_1^2 + \sigma_2^2) + e^{\tilde{f}(r)} \sigma_3^2\right) \,, 
\\
    &
     \equiv
  - \tilde{a}(r) dt^2 + \tilde{b}(r) dr^2 + \frac{1}{4} r^2
\Big( e^{\tilde{c}(r)} d\theta^2 +d\phi^2 \left(e^{\tilde{c}(r)} \cos ^2(\theta )+e^{\tilde{f}(r)} \sin ^2(\theta
    )\right) \nonumber 
\\
    &~~~-2 e^{\tilde{f}(r)} \sin (\theta ) d\psi 
d\phi +e^{\tilde{f}(r)} d\psi^2\Big)
\,,
 \label{19VII25.2}
\end{align}
where we used
\begin{equation}
\sigma_1 + i \sigma_2 = e^{i \psi} \left(\cos \theta \, d \phi + i \,d \theta\right)\,, \qquad \sigma_3 = d \psi - \sin\theta d \phi\,,
\end{equation}
with $0 \leq \theta < \pi$ and $0 \leq \phi, \psi < 2 \pi$.
The Birmingham-Kottler metrics~\cite{Birmingham,Kottler} are a special case of \eqref{19VII25.2}, with 
\begin{equation}\label{19VII25.1}
   \tilde{a}(r) = r^2 +1 -\frac{2m}{r^2} = (\tilde{b}(r))^{-1}
   \,, 
   \quad
   \tilde{c}(r) \equiv 0 \equiv \tilde{f}(r)
   \,.
\end{equation}
%
% \ptc{alternative forms of the metric moved to alternative.tex}
%
%\input{alternative} 

In  regions where the volume of the orbits of  $SO(3)\times U(1)$ has non-vanishing gradient (equivalently, in regions where all orbits have non-zero mean curvature  
on the level sets of $t$)  
we can redefine   $r$ to   a ``volume coordinate'', which corresponds to the requirement
 $\tilde f = -2 
\tilde c$. In order to encode the ALH asymptotics it is 
further convenient to replace the new $r$ by a coordinate $x$ defined by 
\begin{equation}
 r = \tan x
 \,.
 \label{rtox}
\end{equation}
%.   
For the sake of simplicity of the field equations it is further convenient to reparameterise the metric functions as in~\cite{BizonRostworowskiAds5}%
\footnote{Our form of the metric is identical to that considered in~\cite{BizonRostworowskiAds5}  but we are assuming $t$-independence and a different asymptotics at infinity. Indeed, Bizo\'n and Rostworowski consider metrics with a conformal boundary identical to that of AdS$_5$, while our conformal infinity is different.}
 (see \cite{Lipert.2010} and \cite{Bizon:2005cp6w9} for a similar form of the metric in $6+1$ and $8+1$ dimensions):  
\begin{eqnarray}
	\label{eq:25.07.23_01}
	g 
 &= &
   \frac{1}{\cos^{2}\red{(x)}} \Big(
  -\redA(x)  e^{-2\delta(x)}dt^{2} + \redA(x) ^{-1}dx^{2}
   \nonumber
\\
 && 
     + \frac{1}{4}\sin^{2}\red{(x)}\left(e^{2B(x)}\big(\sigma_{1}^{2}
 +\sigma_{2}^{2}\right)+e^{-4B(x)}\sigma_{3}^{2}\big) 
  \Big)
  \,.
\end{eqnarray}
Here  $x\in[0,\pi/2)$ for solutions without event horizons, and $x
\in(x_h,\pi/2)$, with some $x_h>0$ at which $\redA$ vanishes, for black hole solutions.
In this  parameterisation % \eqref{eq:25.07.23_01}  
the vacuum static Einstein equations with cosmological constant $\Lambda=-\blue{6}$
 read%
\footnote{In equation (2) of~\cite{BizonRostworowskiAds5} the first exponent should be $+2B$ and the second one should be $-4B$ in order to reproduce their equations (4)-(7).}
\begin{align}
	\label{eq:25.07.23_02}
	\delta' &= -2\sin\red{(x)}\cos\red{(x)}\left(B'\right)^{2}\,,
	\\
	\label{eq:25.07.23_03} 
	( e^{-\delta} \redA) ' &= e^{-\delta}\big(
 4\tan\red{(x)}(1-\redA ) 
 + \frac{2 \cot(\red{x}))}{3}( 4e^{-2B}-e^{-8B}-3\redA)
 \big)\,,
	\\
	\label{eq:25.07.23_04}
	0 &= \frac{1}{\tan^{3}\red{(x)}}\left(\tan^{3}\red{(x)} \redA  e^{-\delta}B'\right)' - \frac{4e^{-\delta}}{3\sin^{2}\red{(x)}}\left(e^{-2B}-e^{-8B}\right)\,.
\end{align}
The  freedom of adding a constant to $\delta $ can be absorbed into a rescaling of $t$.
%\includegraphics[width=.8\textwidth]{Aprime.png}
%
%\includegraphics[width=.8\textwidth]{mass.png}
%}
\subsection{Asymptotics}
 \label{ss25XII25.2}
 
We seek  solutions that have a finite limit as  $x\to\pi/2$, where the equations have a Fuchsian singularity. A consistent polyhomogeneous expansion is obtained by setting
\begin{align}\label{12VIII25.1a}
  \redA(x)= 
   \ &1 + \bredA_2 \left(x-\frac{\pi}{2}\right)^2 + \bredA_4 \left(x-\frac{\pi}{2}\right)^4
     + \bredA_{\log} \left(x-\frac{\pi}{2}\right)^4 
   \log \big \vert x-\frac{\pi}{2}\big \vert 
   \nonumber 
\\
   &+  \bredA_5 \left(x-\frac{\pi}{2}\right)^5 +
   \ldots
 \,,
\\ \label{12VIII25.1b}
  B(x)= 
   \ &\Binf_0 + \Binf_2 \left(x-\frac{\pi}{2}\right)^2 + \Binf_4 \left(x-\frac{\pi}{2}\right)^4 + \Binf_{\log} \left(x-\frac{\pi}{2}\right)^4 
   \log  \big \vert x-\frac{\pi}{2} \big\vert \nonumber \\
   &+  \Binf_5 \left(x-\frac{\pi}{2}\right)^5+ 
   \ldots
 \,,
\\ \label{12VIII25.1c}
  \delta(x)= 
   \ &  \bdelta_ 4 \left(x-\frac{\pi}{2}\right)^4 +  \bdelta_5 \left(x-\frac{\pi}{2}\right)^5+ 
   \ldots
 \,.
\end{align}
Inserting this in the equations 
one finds
 \ptcignore{consistenly with  the mass formula $A_2$ is negative, with a max at 0, in the range of interest ; note also that the other coefficients have signs for large B}
\begin{align}\label{12VIII25.1} 
     \bredA_2
    =
    \ & 
       \frac{\blue{1} }{3} e^{-8 \Binf_0} \left(4 e^{6 \Binf_0}-1\right)- 1
       \red{\le 0} 
   \,, 
   \\
   \Binf_2
    =
    \ & 
      -\frac{1}{3} e^{-8 \Binf_0} \left(e^{6
   \Binf_0}-1\right)
   \,,
\\
   \Binf_{\log} 
    =
    \ & 
       \frac{2}{9} e^{-16 \Binf_0} \left(-5 e^{6
   \Binf_0}+e^{12 \Binf_0}+4\right)
   \,, 
\\ 
   \bdelta_4 = \ & \frac{2}{9} e^{-16 \Binf_0} \left(e^{6
   \Binf_0}-1\right)^2 
   \red{= 2 \Binf_2^2 \ge 0}\,,
    \label{12VIII25.22a} 
\\
   \label{12VIII25.1log} 
   \bredA_{\log} 
    =
    \ & 
       -\frac{8}{9} e^{-16 \Binf_0} \left(e^{6
   \Binf_0}-1\right)^2
    {\redc = - 8 \Binf_2^2 \le 0 }
   \,.
\end{align}
It follows%
\footnote{As such, Biquard's theorem applies to a metric where $-dt^2$ has been replaced by $dt^2$, and this can be used to justify our assertion.} 
from~\cite{Biquard}
 that all the coefficients in a polyhomogeneous expansion of $g$ are uniquely determined by $\Binf_0$, $\Binf_4$ and $\bredA_4$.  
  A calculation shows that the holographic mass, plotted in Figure~\ref{F13VII25.1} for horizon-less solutions, is determined by $\Binf_0$ and $\bredA_4$.  
  
\subsection{Conformal infinity}
 \label{ss25XII25.1} 
A representative of the conformal class of metrics at infinity is given by
\begin{equation}
	\label{22XII25.2} 
  - dt^{2} + \frac{1}{4} \left(e^{2\Binf_0}\left(\sigma_{1}^{2}
 +\sigma_{2}^{2}\right)+e^{-4\Binf_0}\sigma_{3}^{2}\right)  
  \,.
\end{equation}
The metric induced on the level sets of $t$ by \eqref{22XII25.2} is  Einstein if and only if $\Binf_0=0$, from which it should be clear 
 that $g$ does  \emph{not} belong to the Birmingham-Kottler family~\cite{Birmingham,Kottler} 
 unless  $\Binf_0=0$. 
This last fact follows in any case from \eqref{12VIII25.1a}-\eqref{12VIII25.22a}, as the $(x- {\pi}/{2} )^4 
   \log   \vert x- {\pi}/{2}  \vert$--part of these expansions defines a conformally  covariant tensor field on the conformal boundary which vanishes for metrics conformal to \eqref{22XII25.2} if and only if $\Binf_0=0$, in which case a solution  without horizons is the anti-de Sitter metric;
   see also Appendix~\ref{s13VII26.1}.
    \ptcignore{added without horizons? with horizons should be BK? give ref in any case, there should be one for AdS\\ -- \\ there should be a uniqueness result in this class}
     
Numerics shows that one can find solutions both with or without horizons for any value of $\Binf_0 \in \R$. When $\Binf_0$ tends to negative infinity the boundary looks then like a very long and very thin bicycle tube, and when $\Binf_0 $ tends to infinity the boundary looks like a very big sphere with small circles attached. The Kretschmann scalar of the boundary metric \eqref{22XII25.2}  reads
\begin{equation}
R_{\mu \nu \rho \sigma} R^{\mu \nu \rho \sigma} =44 e^{-16 \Binf_0}-96 e^{-10
   \Binf_0}+64 e^{-4 \Binf_0}\,.
\end{equation}
We see that the scalar $R_{\mu \nu \rho \sigma} R^{\mu \nu \rho \sigma} $
becomes arbitrarily large  when $\Binf_0 \rightarrow - \infty$ and goes to zero as $\Binf_0 \rightarrow \infty$.
The radius of injectivity tends to zero in either limit.

The coefficients  $\bredA_4$ and $\Binf_4$ are defined 
globally by the solution (as opposed to locally by the conformal geometry at infinity), with $\bredA_4$ related to the mass.
%, see Section~\ref{s14VIII25.1} below. 

The limit $B\to\infty$ is considered in Appendix~\ref{ss10VIII25}.

\subsection{Solutions without horizons}
 \label{ss2VIII25.1}
 
For conformally compactifiable  solutions without event horizons, the smoothness at $\{x=0\}$ together with the field equations implies the following expansion,
for small $x$:
\begin{align}
	\label{eq:25.07.23_05}
	\delta &= \blue{\delta_0} -2 B_2^2 x^4 + \frac{4}{3} B_2^2 \left(10 B_2+1\right) x^6 + \red{O}(x^{8})\,,
	\\
	\label{eq:25.07.23_06}
	\redA  &= 1 -4 B_2^2 x^4 + \frac{1}{3} B_2^2 \left(80 B_2+11\right) x^6 + \red{O}(x^{8})\,,
	\\
	\label{eq:25.07.23_07}
	B &= B_2 x^2 -\frac{1}{12} B_2 \left(30 B_2+1\right) x^4 + \frac{1}{90} B_2 \left(810 B_2^2+15 B_2-2\right) x^6+ \red{O}(x^{8})\,,
\end{align}
with  constants  $\blue{\delta_0},B_2\in \R$. 

Numerics
shows
 that spatially-conformally compact ALH vacuum solutions  without event horizons exist for 
$$
 B_{2*} \approx -0.1046384<B_{2}<
 \infty
 \,.
$$
%.
Note that existence of solutions for small $|B_2|$ follows from the results quoted in the Introduction. An existence proof throughout the range could be carried out by numerically bounding the spectrum of the Lichnerowicz operator away from zero, but we have not attempted to do this. 

\begin{figure}
	\centering
 	\includegraphics[width=0.32\textwidth]{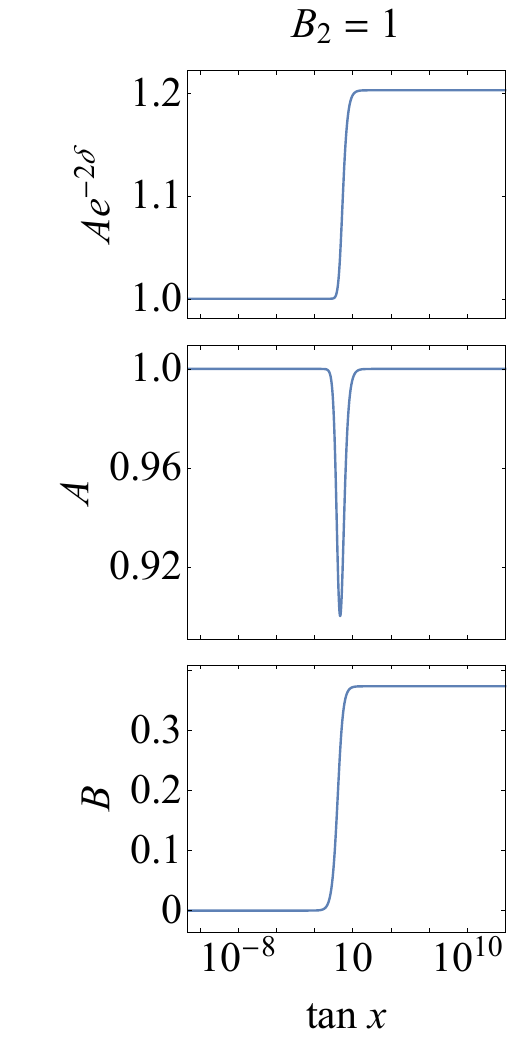}
 	\includegraphics[width=0.32\textwidth]{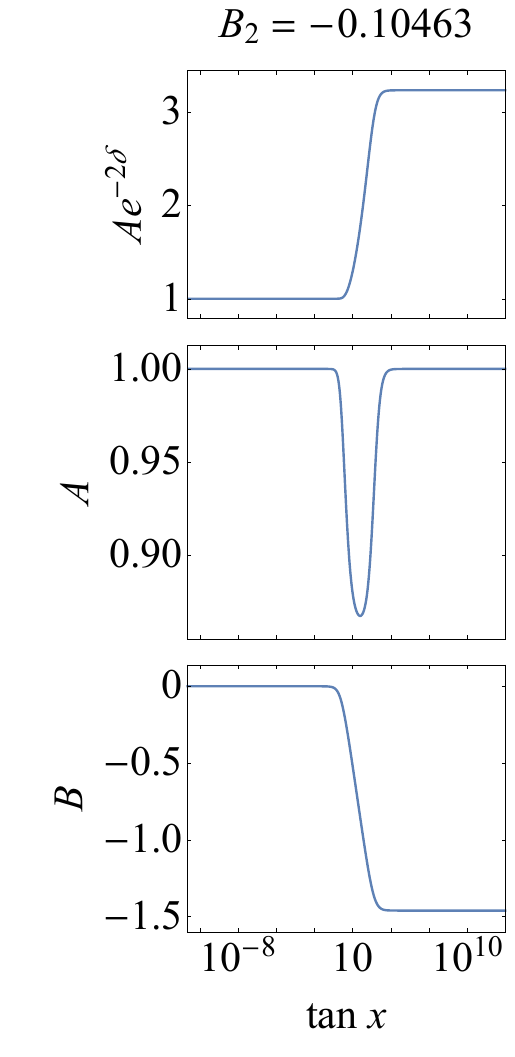}
 	\includegraphics[width=0.32\textwidth]{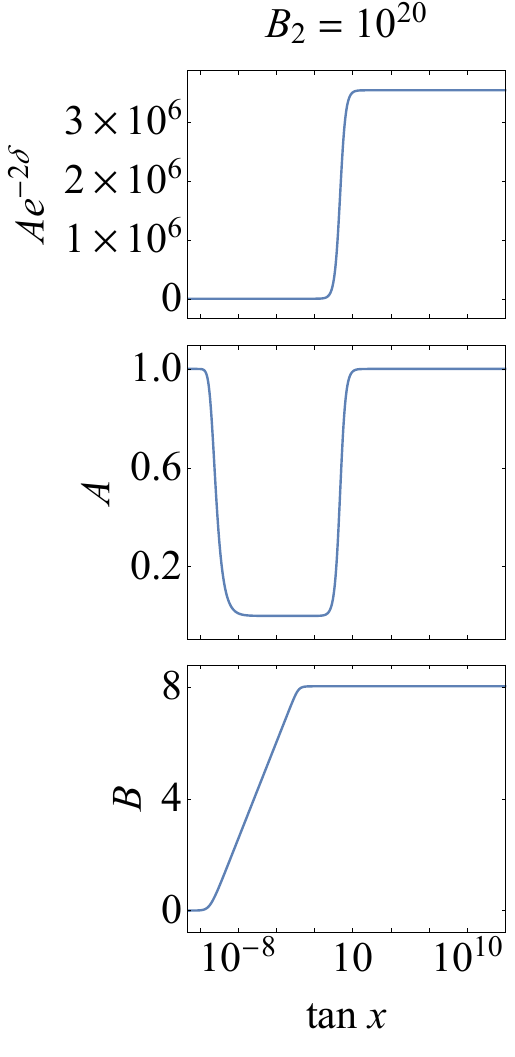}
 	\caption{Plot of the metric functions $Ae^{-2\delta}$, $\redA$ and $B$ 
 as functions of $\tan x$, with logarithmically scaled horizontal axis, for a sample of solutions of \eqref{eq:25.07.23_02}-\eqref{eq:25.07.23_04} without event horizons, with $B_{2}\gtrsim B_{2*}$ (left row), $B_{2}=1$ (middle row), and  $B_{2}=10^{20}$ (right row).}
\label{F2VII25.1} 
\end{figure}

Representative plots of the functions $\redA $, $B$ and $A e^{-2\delta}$ are found in Figure~\ref{F2VII25.1}. 
 All three functions tend to finite limits at the conformal boundary at infinity $\{x=\pi/2\}$, consistently with the requirement of existence of a conformal completion.  

\begin{figure}
	\centering
 	\includegraphics{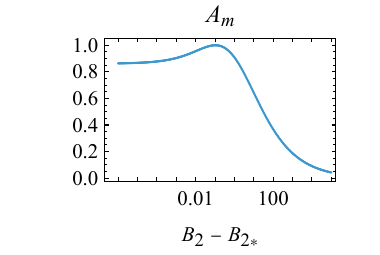}
 	\includegraphics{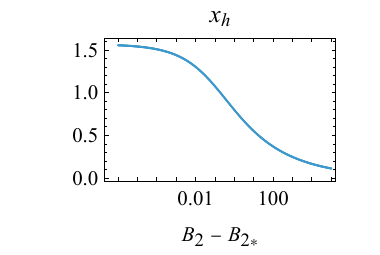}
	\caption{Plot of  $A_m:=\min A$, and of the value $x_h$ of $x$ for which $\min A $ is attained, as functions of  {$B_2 - B_{2*}$}. 
}
\label{F2VII25.21} 
\end{figure}

It is of interest to enquire what breaks down when the limits $B_2\searrow B_{2*}$ and $B_2\to \infty$ are approached. Since  
black-hole solutions  are characterised by the existence of $x_h$ such that $\redA (x_h)=0$, the property that $\redA$ tends to zero could be interpreted as evidence of black hole formation.   
Figure~\ref{F2VII25.21} hints  strongly at  $\lim_{B_2\to\infty}\min \redA=0$, suggesting indeed the formation of a strongly distorted black hole; compare the discussion of the limiting geometry in Section~\ref{ss10VIII25}.  

\begin{figure}
	\centering
	 	\includegraphics{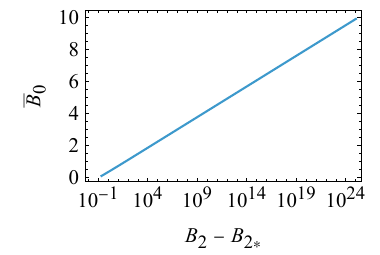} 
 	\includegraphics{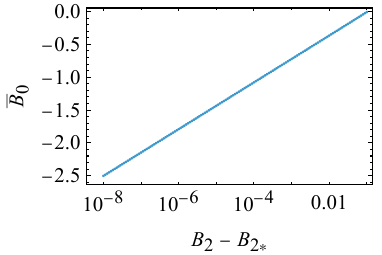} 
 	\caption{Plot of $\Binf_0$ for large $B_2$ (left) and for small $B_2- B_{2*}$ (right).}
\label{F10V26.1} 
\end{figure}

We find that $\Binf_0$ increases monotonically with $B_2$. The graphs of  Figure~\ref{F10V26.1} suggest strongly that
\begin{equation}\label{14VII26}
  \lim_{B_2\to\infty} \Binf_0 = \infty
  \,,
  \qquad
  \lim_{B_2  \searrow B_{2*}} \Binf_0 = - \infty
  \,.
\end{equation}
Indeed, the graph of the left Figure~\ref{F10V26.1}  is very well approximated by  
%	large: 
\begin{equation}\label{10V26.1y}
 \Binf_0 \approx    0.342832 + 0.166666 \log( B_2+0 .104638 )
\end{equation}	  
for large $B_2$.
The graph of the right Figure~\ref{F10V26.1} is very well approximated by  
%	large: 
\begin{equation}\label{10V26.1y2}
 \Binf_0 \approx    {0.346721} +  {0.154723}  \log( B_2  - B_{2* }  )
\end{equation}	  
for small  $ B_2-B_{2* } > 0 $. 
%

    %\section{Kretschmann scalar}

\begin{figure}
	\centering
	\includegraphics{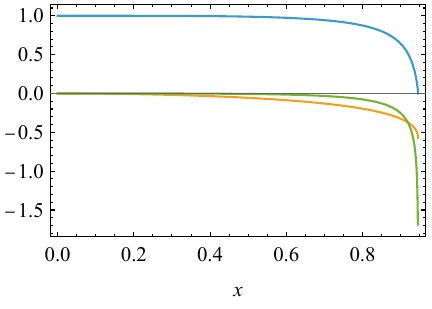} 
	\includegraphics{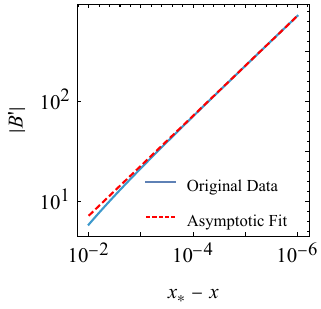} 
 	\caption{Plots of the metric functions $ \delta $ (green), $\redA$ (blue), $B$ (orange) (left figure), and $  |B'| $ (right figure), with    $B_2=-0 .2 < B_{2*}$.}
\label{F10V26.2} 
\end{figure}

For $B_2<B_{2*}$ the solution stops with $B'$ tending to minus infinity for a finite value of $x$, say $x=x_*$.
 A representative plot can be found in   Figure~\ref{F10V26.2}.   
Numerics suggests that this is a curvature singularity. We note that, after using the field equations,  the Kretschmann scalar of  the metric \eqref{eq:25.07.23_01} 
can be written as a polynomial of order 6 in  $B'(x)  $, with coefficients which depend upon $A(x)$ and  $B(x)$ but are  independent of $\delta$, with
% \ptc{ the 5 has asquare, the 4 and 3 have an a, and 2 1 0 have no multiplicative a}
% 
\begin{equation}
R_{\mu \nu \rho \sigma} R^{\mu \nu \rho \sigma}
=  
    48  
   \sin^2 (x) \cos^6 (x)
   {\redA(x)^2} B'(x)^6 + ... 
   \,. 
    \label{12VI26.1}
\end{equation} 
Numerics suggests that when  $B_2 \nearrow B_{2*}$  the scalar
$R_{\mu \nu \rho \sigma} R^{\mu \nu \rho \sigma}$  blows up as $(x_*-x) ^{-\alpha}$, 
with $\alpha$ close to $1$ , as the boundary $x_*$ of the domain of existence of the solution is approached, with $\delta$ tending to minus infinity,  $A$ approaching zero, and  $B$ approaching a finite value. The numerically observed limit $\lim_{x\to x_*} A(x)=0$ does not seem to compensate for the blow-up of $B'$ in \eqref{12VI26.1}.   
Numerics  also indicates that when $B_2$ tends to $-\infty$,  the blow-up point $x_*$ tends to zero, with  
$$
 \lim_{B_2\to-\infty} \lim_{x\to x_*}B  \approx -.46
\,.
$$

\subsection{Black holes}
 \label{ss2VIII25.2}

As already mentioned, 
black-hole solutions  are characterised by the existence of $x_h\in (0,\pi/2)$ such that $\redA (x_h)=0$. 
A first-order zero  at $x_h$ corresponds to bifurcate horizons with non-vanishing surface gravity $\kappa$, a zero of higher order would correspond to a degenerate, i.e.\ $\kappa=0$, horizon (cf., e.g., \cite[Sections~6.3.1-2]{Chrusciel:2020fql}). 
The field equations
  and smoothness at the totally geodesic submanifold $\{x=x_h\}$ imply the following Taylor expansion
\begin{align}
	\label{eq:25.07.23_08}
	\delta &= \bdelta_h-\frac{8 \left(x-x_h\right) \left(\left(e^{6 B_h}-1\right){}^2 \cot
   ^3\left(x_h\right) \csc ^2\left(x_h\right)\right)}{\left(\left(4 e^{6
   B_h}-1\right) \cot ^2\left(x_h\right)+6 e^{8
   B_h}\right){}^2}+O\left(\left(x-x_h\right){}^2\right)
    \,,
	\\
	\label{eq:25.07.23_09}
	\redA  &= \left(x-x_h\right) \left(\frac{2}{3} e^{-8 B_h} \left(4 e^{6 B_h}-1\right)
   \cot \left(x_h\right)+4 \tan
   \left(x_h\right)\right)+O\left(\left(x-x_h\right){}^2\right)
    \,,
	\\
	\label{eq:25.07.23_10}
	B &= B_h+\frac{2 \left(x-x_h\right) \left(e^{6 B_h}-1\right) \cot \left(x_h\right)
   \csc ^2\left(x_h\right)}{\left(4 e^{6 B_h}-1\right) \cot
   ^2\left(x_h\right)+6 e^{8 B_h}}+O\left(\left(x-x_h\right){}^2\right)
   \,,
\end{align} 
for a constant $B_h\in \R$.  
As already mentioned, the constant $\bdelta_h$ can be absorbed into a redefinition of the coordinate $t$, and therefore without loss of generality one can assume that, e.g.,
\begin{equation}\label{9V26.1}
  \bdelta_h = 0
  \,.
\end{equation}

\begin{figure}
	\centering
%	\includegraphics[width=0.54\linewidth]{existence_bh.pdf}
%    \hfill % reproduced Figure 2.5
	\includegraphics[width=0.45\linewidth]{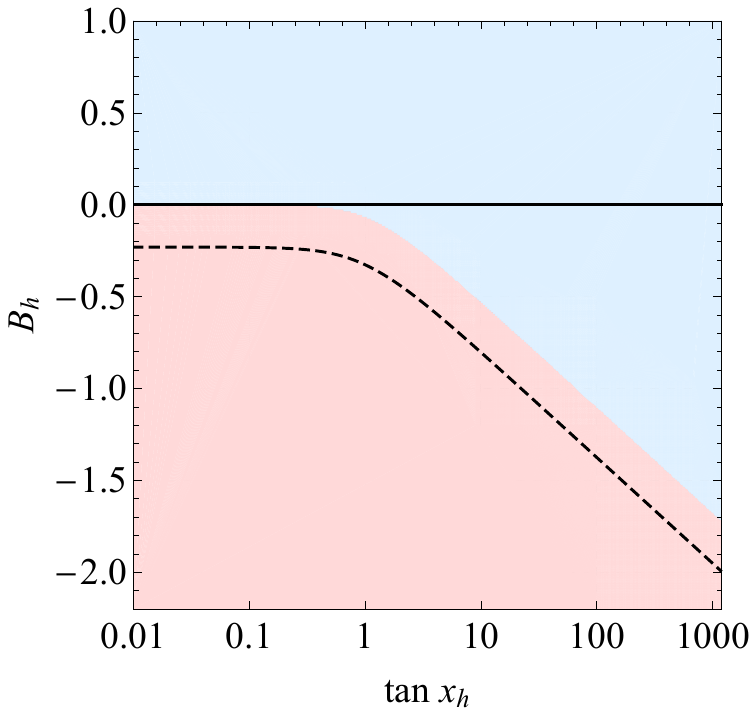}
	\caption{The boundary of the existence of black hole solutions on the $(x_h, B_h)$ plane. Black holes  exist in the blue region. 
The Birmingham-Kottler black holes  lie on the line $B_h=0$, with the boundary lying below this line.  The  dashed line indicates the location where the surface gravity would vanish, making clear that degenerate black holes do not occur in this family. }
	\label{F2VII25.4} 
\end{figure}

We find that the  $(x_h,B_h)$-plane  splits into two connected regions, where for each  $x_h>0$ black holes exist for all $B_h$ above a threshold $B_h>B_*(x_h)$, see Figure~\ref{F2VII25.4}. It is not apparent from the figure but we have $B_*(x_h)<0$ for $x_h<0$, with $B_*(x_h)\to 0$ as $x_h\to 0$; recall that the line $B_h=0$ corresponds to Birmingham-Kottler black holes. 

\begin{figure}
	\centering  
	\includegraphics[width=.315\textwidth]{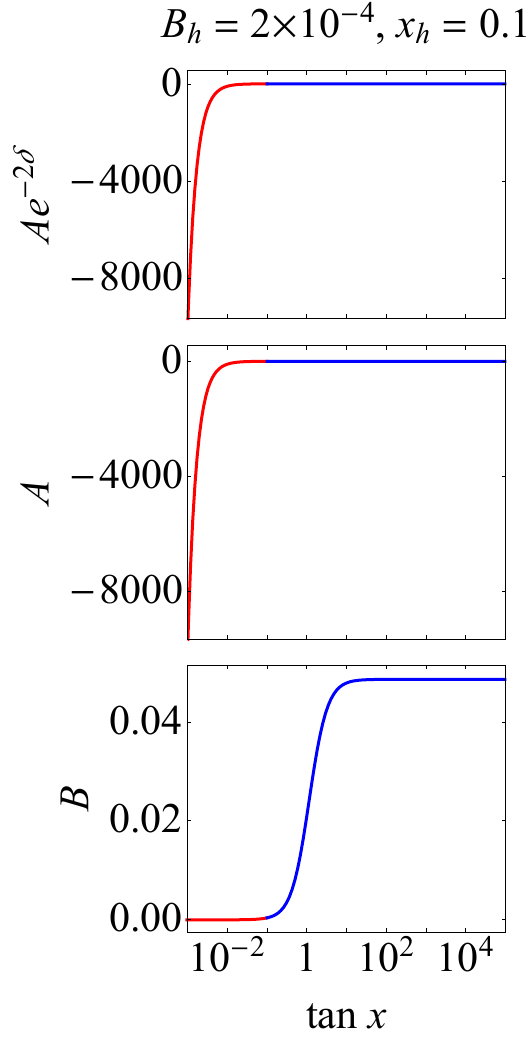} 
	\includegraphics[width=.315\textwidth]{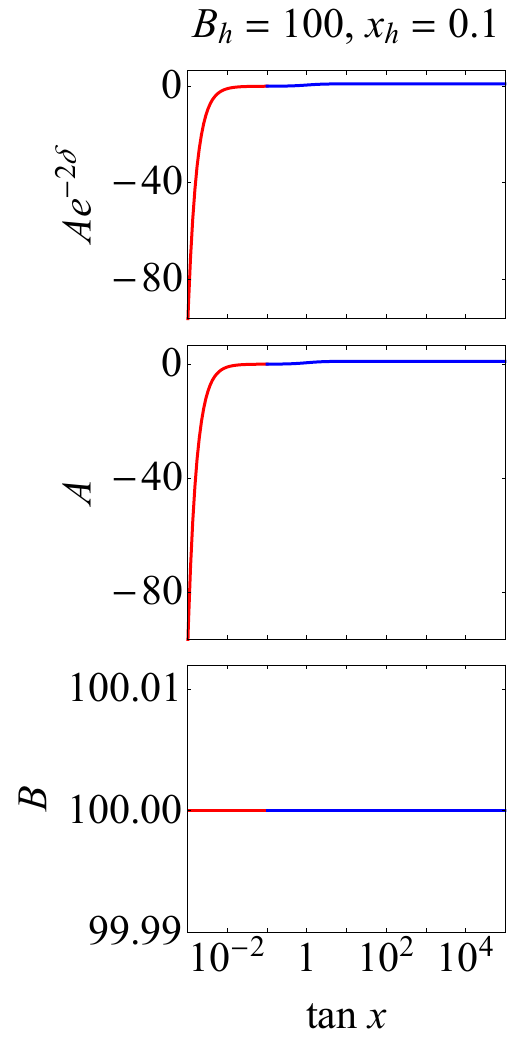}
	\includegraphics[width=.351\textwidth]{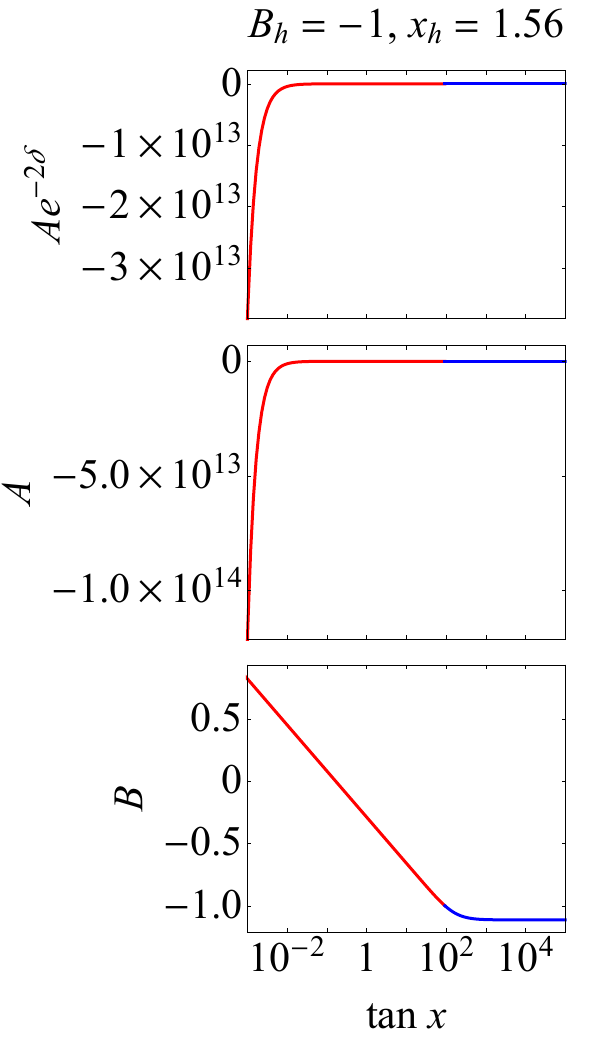}
	\caption{Example plots of the metric functions $A e^{-2\delta}$, $\redA $ and $B$ for black-hole solutions  of \eqref{eq:25.07.23_02}-\eqref{eq:25.07.23_04}. The blue curve is the part of the solution above the Killing horizon. 
}
\label{F2VII25.3} 
\end{figure}

Three representative plots of the metric functions can be found in Figure~\ref{F2VII25.3}.

\begin{figure}
	\centering
	\includegraphics[width=.6\textwidth]{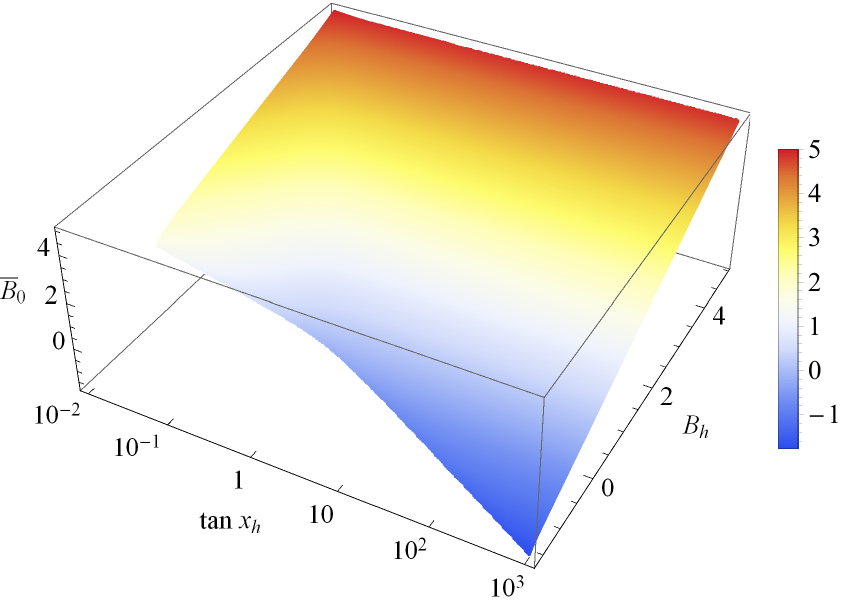}
	\caption{The asymptotic value $\Binf_0$ of $B$ as a function of $B_h$ and $x_h$.}
\label{F2VII25.4b} 
\end{figure}

 The asymptotic value $\Binf_0$ of $B$ is plotted in Figure~\ref{F2VII25.4b}.  
We find that $\Binf_0\approx B_h$ when either $\tan(x_h)$ or $ B_h $ are large, which provides evidence that arbitrarily large values of $\Binf_0$ are attained:
$$
    \lim_{B_h\to \infty}\Binf_0 =  \infty
 \,.
$$
%.
(See also Appendix~\ref{ss10VIII25}.) 
The   boundary curve in Figure~\ref{F2VII25.4} is well approximated by 
\begin{equation}
 \label{21VII26.1}
 B_h\approx -0.573811 \log_{10}\big(\tan (x_h)\big)+0.051013
\end{equation}
for large $\tan (x_h)$. We further find that $\Binf_0 < B_h$ along the boundary curve for large $\tan (x_h)$, which provides evidence that arbitrarily large negative values of $\Binf_0$ are achieved. 
  
 In view of the Birmingham-Kottler metrics \eqref{19VII25.1}, for every asymptotic value $\Binf_0$ within the allowed range one expects a one-parameter family of black hole solutions with different masses.

After extending the metric across $\mcH:=\{x=x_h\}$ in the 
usual way, the hypersurface $\mcH$ becomes a Killing horizon in the extended spacetime, and for any constant $k\in\R$  the vector field $\xi = k \partial_t$ is null at    $x = x_h$.
The surface gravity can be calculated as in \cite[Section 4.6.3]{Chrusciel:2020fql}, 
\begin{equation}
\kappa = - \frac{1}{2} b^\mu \nabla_\mu (\xi^\nu \xi_\nu) \vert_{x = x_h}\,,
\end{equation}
where $b$ is any one-form extending smoothly across the horizon which satisfies $b(\xi) =1$. Choosing 
$$
 b= \frac{1}{k} \left( dt + \frac{dx}{A(x) \exp(-\delta(x))} \right) =: dv
  \,,
$$
%, 
where $v$ is smooth across the horizon in the extended manifold,
we find  
\begin{align}
\kappa &= \frac{1}{2} k e^{-\delta (x)}
   \left(A'(x)+2 A(x) \left(\tan
   (x)-\delta '(x)\right)\right) \vert_{x = x_h } 
\\
   &= \frac{1}{2} e^{-\bdelta_h} k
   \left(\frac{2}{3} e^{-8 B_h}
   \left(4 e^{6 B_h}-1\right) \cot
   (x_h)+4 \tan
   (x_h)\right)\,.
\end{align}

Recall  that $(\phi,\psi,\theta)$ are  Euler angles parameterising  the rotation group which is doubly covered by the sphere. We have the freedom to decide whether the horizon topology is that of the rotation group or of the sphere. Assuming the former, the entropy of the horizon equals
\begin{equation}
S = \frac{A_h}{4} = 
\frac{1}{32}
\int_{  SO(3) } \tan^3(x_h)  \,
|\cos\theta| \,
 d\psi \, d \theta  \,
 d \phi
= 
\frac{\pi^2}{ 4} \tan^3(x_h)\,,
\end{equation}
where $A_h$ is the area of the horizon. %
%%  
%\begin{figure}
%	\centering
%	\includegraphics{Bpi2_vs_BB2.pdf}
%	\caption{Plot of $B(\pi/2)$, $\delta(\pi/2)$, $\redA (\pi/2)$ as functions of $B_{h}\in ???$ for black hole solutions. MM: todo}
%\label{F2VII25.4a} 
%\end{figure}
%%

It has been shown in~\cite{CRT} that the metrics induced on compact sections of event horizons of static vacuum black holes with $\kappa=0$ are Einstein.  
       In our Ansatz this would require $B_h=0$,  whence 
       $$
\kappa=
e^{-\bdelta_h} k  \left(\cot (x_h)+2 \tan (x_h)\right)
 \,,
   $$
   which never vanishes: there are no degenerate solutions in our family of metrics.

\ptcignore{alternative analysis in alternative2.tex}

\bigskip
\red{
    \noindent{\sc Acknowledgements:}
RW acknowledges
support from the STFC consolidated grant ST/X000583/1 “New Frontiers in Particle Physics,
Cosmology and Gravity”. 
}

\appendix

 \section{$B\equiv 0$}
 \label{s13VII26.1} 

When $B\equiv 0$ the vacuum Einstein equations \eqref{eq:25.07.23_02}-\eqref{eq:25.07.23_04} reduce to 
\begin{align}
	\label{6VI26.p1}
	\delta' &=0\,,
	\\
	\label{6VI26.p2} 
	  \redA  ' &=  
2 \big( 2\tan\red{(x) {+} \cot(\red{x}) \big)}  (1-\redA )  
 \,, 
\end{align}
which are the Birmingham-Kottler (``Schwarzschild-AdS'') black holes in disguise. Indeed, rescaling $t$ so that $\delta \equiv0$, keeping in mind that $ r = \tan x$,  the metric reads  
\begin{eqnarray}
	\label{6VI26.p3} 
	g 
 &= &
   \frac{1}{\cos^{2}\red{(x)}} \Big(
  -\redA(x)  dt^{2} + \redA(x) ^{-1}dx^{2}
  \Big)
   \nonumber
\\
 &&   
     +r^2 \underbrace{ \frac{1}{4}
    (d \theta^2 + d \phi^2 + d \psi^2 -2 d \phi d\psi \sin\theta)}_{d\Omega^2}
  \,,
\end{eqnarray}
where $d\Omega^2$ is the unit round metric on \blue{$S^3$}, with
\begin{equation}\label{6VI26.p4}
  \frac{A(x)}{\cos^2(x)} = \big(
   r^2 +1 - \frac{2m}{r^2}
   \big)\big|_{r=\tan x}
    = 
  \tan^2(x)  +1 -  2m \cot^2(x)
  \,;
\end{equation}
equivalently 
\begin{equation}\label{6VI26.p5}
  A(x)  =  
 1 - 2m \frac{\cos^4(x)}{\sin^2(x)}
 \,.
\end{equation}
We see that  
the Birmingham-Kottler mass parameter $m$
 is determined by $x_h$ as 
\begin{equation}\label{6VI26.p6}
  A(x_h)  =  0 
  \qquad
  \Longleftrightarrow
  \qquad
   m = 
   \frac{\sin^2(x_h)}{2\cos^4(x_h)}
 \,.
\end{equation}

\section{Large $B$ limit}
 \label{ss10VIII25}
 
A very large constant $B$ provides an approximate solution of \eqref{eq:25.07.23_04}. Equation~\eqref{eq:25.07.23_02} implies then that $\delta$  is approximately constant, while in the limit $B\to\infty$ Equation \eqref{eq:25.07.23_03} becomes
\begin{align} 
	\label{eq:25.07.23_03bc}
	\redA ' &= 4\tan{x}(1-\redA ) -  \frac{2\redA }{\tan{x}}
 \,.
\end{align}
All solutions on $(0, \pi/2)$ are given by
\begin{equation} 
 	\bredA_c(x) = 1-c\cos^{2}{x} + (c-1)\cot^{2}{x}\,,
    \label{16VIII25.24}
\end{equation}
where $c\in\R$ is a constant. 
Regularity at $x=0$ requires $c=1$, in which case one obtains  
\begin{equation}
    \label{A1}
 \bredA_{1}(x) = \sin^{2}{x}
 \,.
\end{equation}
%, 
Numerics shows that $\redA(x)$ approaches  $\bredA_{1}(x)$
as the parameter $B_2$ in \eqref{eq:25.07.23_05}-\eqref{eq:25.07.23_07} below tends to infinity.

As such, the  functions $\bredA_c$  provide  solutions of the vacuum Einstein equations after appropriate scalings of the coordinates. Namely, to compensate for the large expansion of the Hopf spheres of constant $\psi$, and for the shrinking of the Hopf circles of constant $(\theta,
\phi)$ one can redefine the coordinates as
\begin{equation}\label{16VIII25.21}
  \psi \to \bar \psi = e^{2\scaleB }\psi
  \,,
  \quad
  \phi \to \bar  \phi= e^{-\scaleB }\phi
  \,,
  \quad
  \theta \to \bar  \theta= e^{-\scaleB }\theta
  \,,
\end{equation}
where $\scaleB$ is a constant. Letting $\scaleB$ tend to infinity in  \eqref{eq:25.07.23_01} one obtains the metrics
\begin{equation}
	\label{eq:25.07.23_01cd}
	g = \frac{1}{\cos^{2}{x}} \left(
  -\redA  e^{-2\delta}dt^{2} + \redA ^{-1}dx^{2} + \frac{1}{4}\sin^{2}{x}
   \left(d\bar\psi^2 + d\bar\theta^2 + d\bar\phi^2
   \right)
  \right)
  \,.
\end{equation}
When $\delta$ is a constant the   metrics \eqref{eq:25.07.23_01cd} 
 with $A = \bredA_{c} $
are Birmingham-Kottler metrics with toroidal infinity and with mass parameter $m=(1-c)/32$. This can be seen by setting $r= \tan(x)/2$ and $t = \bar{t} \exp(\delta)\blue{/2}$,  which leads to the standard form of Birmingham-Kottler metrics in spacetime dimension five:
\begin{eqnarray}
g 
 & = &
 -( r^2 - 2m/r^2)   d\bar{t}^2 + \frac{dr^2}{ r^2 - 2m/r^2} +  r^2 ( d\bar\psi^2 + d\bar\theta^2 + d\bar\phi^2 
  )
%  \nonumber
%\\
% & = &
%  \bar y ^{-2} \big(\blue{-}d\bar{t}^2 + d\bar y^2 +  d\bar\psi^2 + d\bar\theta^2 + d\bar\phi^2    
%  \big)
    \,.
\end{eqnarray}

\ptcignore{the bibliography has to be uniformised eventually, abbreviated names of journals and first names of authors, arXiv ref whenever available as a note \\ -- \\ rw: done}

\bibliographystyle{amsplain}

\bibliography{numsol-minimal,ChruscielWutte-minimalkmunu}

\end{document}